\documentclass[letterpaper]{JHEP}
\usepackage{epsfig}
\usepackage{amsmath}
\usepackage{graphicx}
\usepackage{amsfonts}
\usepackage{amssymb}
\newcommand{\beq}{\begin{equation}}
\newcommand{\eeq}{\end{equation}}
\newcommand{\beqa}{\begin{eqnarray}}
\newcommand{\eeqa}{\end{eqnarray}}

\def\p{\partial}

\def\ie{{\it i.e.,}\ }
\def\eg{{\it e.g.,}\ }

\let\bar=\overbar

\def\Dslash{\not{\hbox{\kern-4pt $D$}}}
\def\dslash{\not{\hbox{\kern-2pt $\del$}}}
\def\p4n{{\mbox{\scriptsize (4+N)}}}
\def\p4{{\mbox{\scriptsize (4)}}}

\def\msb{{\bar{\ssstyle M \kern -1pt S}}}

\newcommand\fverb{\setbox\pippobox=\hbox\bgroup\verb}
\newcommand\fverbdo{\egroup\medskip\noindent%
			\fbox{\unhbox\pippobox}\ }
\newcommand\fverbit{\egroup\item[\fbox{\unhbox\pippobox}]}
\newbox\pippobox
\preprint{hep-th/0203216}
\title{Locally wrapped D-branes}
\author{Fr{\'{e}}d\'{e}ric Leblond$^{a}$, Luc Marleau$^{b}$\\$^{a}$ Department of Physics, McGill University, Montr\'{e}al, Qu\'{e}bec, H3A
2T8, Canada\\$^{b}$ D{\'{e}}partement de Physique, Universit\'{e} Laval, Qu\'{e}bec,
Qu\'{e}bec, G1K 7P4, Canada\\
}
\date{March, 2002\\
March, 2002}

\abstract{We find examples of locally wrapped D-branes in string theory. These excitations
mimic skyrmions in that they correspond to topological excitations of the
scalar fields parametrizing the brane motion in the space transverse to its 
world-volume. While these
brane excitations appear to be point-like, evidence is provided
that curvature corrections to the probe action might allow
for a delocalization of the wrapping on a scale of the order of the string length, 
therefore rendering the phenomena non-singular.}


\begin{document}

\section{Introduction}
Dirichlet branes play a crucial role in string theory. Along
with the fundamental strings, they are the probes with which one must study stringy
backgrounds in order to discover the true nature of quantum geometry. It is therefore 
important to understand the ways in which D-branes can be excited
when they are placed in the curved backgrounds generated by string theory objects. In this work,
we study a generic manner in which branes can be excited when placed in geometries
possessing a compact manifold transverse to their world-volume. 
The main requirement will be that there exists a non-trivial homotopy
mapping for the spatial dimensions of the probe world-volume onto this compact manifold. The
excitations we find correspond to the probe world-volume being wrapped, at a single
point, onto the transverse manifold. The resulting system consists of two orthogonal branes 
intersecting at one point but we give evidence that curvature 
corrections to the probe action might allow for the wrapping to become delocalized
on a scale typical of the string length. Locally wrapped branes mimic skyrmions\cite{skyrme1,skyrme2} 
as they are collective excitations of the open string scalar fields parametrizing the D-brane motion in the space
transverse to its world-volume. 

The paper is organized as follows: In section \ref{description} we give a general description
of locally wrapped D-branes based on the Dirac-Born-Infeld ({\small DBI}) action. We numerically find
an expression for the energy of the excitations. In section \ref{evidence} we consider a typical
$\alpha'$ curvature correction obtained from bosonic string theory and show how it leads to 
a delocalization of the wrapping. In section \ref{applications} we describe 
both supersymmetric and non-supersymmetric
examples in string and M-theory where brane probes can be locally wrapped. We conclude
by discussing to some extent the contribution of the $\alpha'$ curvature
corrections as well as an extension of our approach to more general backgrounds. 
Finally, an appendix introduces a scheme allowing us to analytically describe locally wrapped D3-branes 
approximately, using the Atiyah-Manton ansatz. 

\section{Locally wrapped D-branes}

\label{description}

We refer to a `locally wrapped brane' ({\small LWB}) as a collective excitation of open
string degrees of freedom associated with the world-volume of a D-brane. These
degrees of freedom are the scalar fields characterizing the motion of the
brane in directions that are transverse to its world-volume. The topological
brane excitations we are studying are analogous to branes that
are locally protruding in directions that are transverse to their world-volume by wrapping a compact manifold.
The end result is basically two branes intersecting at right
angles. In string and M-theory, one often encounters so-called wrapped D- or
M-branes (see, \eg ref.~\cite{bachas2}). These are not the same as the states we are introducing in this
paper. Wrapped branes have some or all of their directions wrapped on a
compact manifold. Here, we consider a different kind of wrapping mechanism
where only a finite region of the world-volume is effectively being wrapped.
The world-volume outside of this region is unaffected.

Our approach to studying the {\small LWB}'s is based on the action of a D-brane probe,
the Dirac-Born-Infeld ({\small DBI}) action, in some gravitational background of the
form\footnote{To be consistent we should include a flux associated with the ($p+1$)-form field but
it does not affect the soliton configurations we are considering here.}
\begin{equation}
ds^{2}=G_{\mu\nu}dx^{\mu}dx^{\nu}=\frac{1}{H(\rho)^{k_{1}}}\left[
-dt^{2}+dx^{m}dx^{m}\right]  +H(\rho)^{k_{2}}\left[  d\rho^{2}+\rho^{2}%
d\Omega_{d-p-2}^{2}\right]  , \label{metg}
\end{equation}
\begin{equation}
e^{\Phi}=\frac{1}{H(\rho)^{k_{3}}}, \label{dilg}
\end{equation}
where $m=1,...,p$ so that $d=p+q+2$ and the isometry group is $SO(p,1)\times
SO(d-p-1)$. The metric is here supplemented with a dilaton field, $\Phi$. The form of
the metric and the expression for the dilaton are inspired by the fields sourced
by supergravity $p$-branes. The exponents $k_{1}$, $k_{2}$ and $k_{3}$ are taken as arbitrary
values for now as this will be useful later on when we enlarge the scope to a
more general analysis. We consider placing a D$q$-brane probe on the
background corresponding to eqs.~(\ref{metg}) and (\ref{dilg}). We also use up
($q+1$) gauge degrees of freedom to align the corresponding ($q+1$%
)-dimensional world-volume of the probe parallel to the $t-x^{m}$ coordinates of the
background geometry. The brane is then point-like in the transverse space
parametrized by the coordinates $\rho$ and $\Omega_{d-p-2}$. As pointed out
above, the brane probe motion in the background is, to a good approximation,
governed by the {\small DBI} action\cite{callan1,callan2,leigh},
\begin{equation}
S_{DBI}=-T_{q}\int d^{q+1}\sigma\;e^{-\Phi}\sqrt{-|P[G+B]_{ab}+2\pi
\alpha^{\prime}F_{ab}|}, \label{bi}
\end{equation}
where $a,b=0,...,q$. 
$G_{\mu\nu}$, $B_{\mu\nu}$ ($\mu,\,\nu=0,1,...,d-1$) and $\Phi$ are the
background fields coming from the NS-NS sector of the string
theory. They are respectively the graviton, the Kalb-Ramond and the dilaton
fields. $F_{ab}$ is a U(1) gauge field associated with open string
world-volume excitations and $P[...]$ represents the pull-back operation.
The probe action contains the coupling terms between the induced background fields 
$G_{\mu\nu}$, $B_{\mu\nu}$, $\Phi$ to the open string world-volume excitations. 
As well as the U(1) field, the brane excitations include scalar fields parametrizing their
motion in directions transverse to the world-volume. 
The {\small DBI}
action must be supplemented with the Chern-Simons action including
couplings of the world-volume to fields from the R-R sector. This part of
the action is not playing a role in the upcoming analysis (see section \ref{stringex}
for details). The action eq.~(\ref{bi}) is valid to all orders
in $\alpha^{\prime}$ for couplings involving $F_{ab}$ when the derivatives of field
strenght vanish.

In this work we look at the case where both $B_{ab}$ and $F_{ab}$ are zero.
Studying the effect of these fields might lead to interesting results and
represents a task in itself so we leave that for future investigations. The
simplified version of the {\small DBI} action is then
\begin{equation}
S_{DBI}=-T_{q}\int d^{q+1}\sigma\;e^{-\Phi}\sqrt{-|P[G]_{ab}|}. \label{bi2}
\end{equation}
In section \ref{evidence} we consider how $\alpha^{\prime}$ curvature corrections 
to this action might affect the excitations we now present. Let us consider, for
now, placing a D$q$-brane probe at fixed $\rho$ in a background of the form (\ref{metg}) with
dilaton field given by eq.~(\ref{dilg})\footnote{This assumption is, at this point, unrealistic
since only in certain cases will there be an exact cancellation between the forces exerted on
the D-brane probe by the graviton, dilaton and R-R form field. We will consider such stable systems
in section \ref{applications}.}. The expression for the induced metric
on the probe is
\begin{equation}
P[G]_{ab}=G_{\mu\nu}\frac{\partial X^{\mu}}{\partial\sigma^{a}}\frac{\partial
X^{\nu}}{\partial\sigma^{b}}=G_{ab}+G_{i(a}\partial_{b)}X^{i}+G_{ij}%
\partial_{a}X^{i}\partial_{b}X^{j}, \label{induced}
\end{equation}
where the ($d-p-2$) scalar fields $X^{i}$ ($i=p+2,\,...\,,d-1$) are parametrizing
the brane motion on the compact space $\mathbf{S}^{d-p-2}$ associated with the
factor $d\Omega_{d-p-2}^{2}$ in the metric eq.~(\ref{metg}). Since the probe is
assumed to be fixed at a finite $\rho$, we take $\partial_{a}X^{p+1}=\partial_{a}\rho=0$
\footnote{In section \ref{discussion} we comment on the implications that relaxing this
condition will have on the locally wrapped branes.}. 
For the cases we are interested in, $G_{\mu\nu}$
has no off-diagonal terms so the pull-back of the gravitational field on the
world-volume is simply
\begin{equation}
P[G]_{ab}=G_{ab}+G_{ij}\partial_{a}X^{i}\partial_{b}X^{j}. \label{induced}
\end{equation}
It is useful for our purposes to write down the background metric
eq.~(\ref{metg}) in the form
\begin{equation}
\label{metg2}
ds^{2}=\frac{1}{H(\rho)^{k_{1}}}\left[  -dt^{2}+dr^{2}+r^{2}d\Omega_{p-1}%
^{2}\right]  +H(\rho)^{k_{2}}\left[  d\rho^{2}+\rho^{2}\left(  d\xi^{2}%
+\sin^{2}\xi d\bar{\Omega}_{d-p-3}^{2}\right)  \right]  ,
\end{equation}
where $0\leq\xi\leq\pi$. The first term in this metric is simply a conformally
flat ($p+1$)-dimensional factor written in spherical coordinates while the
transverse part can be regarded as a line (parametrized by $\rho$)
times a sphere $\mathbf{S}^{d-p-2}$ with its volume $V_{\mathbf{S}^{d-p-2}}$
computed with the rescaled radius
\begin{equation}
R=\rho H(\rho)^{\frac{k_{2}}{2}}. \label{radius}
\end{equation}
The $q$-dimensional space-like part of the probe world-volume can be viewed
as being composed of the points on $\mathbf{R}^{q}$ plus a point corresponding
to the `boundary' at spatial infinity. It is then possible to compactify this
infinite hyperplane plus a point on a sphere $\mathbf{S}^{q}$ of infinite
radius. This is a correct idealization of a $q$-dimensional plane as the
curvature of an infinitely large sphere vanishes exactly. The brane
excitations we introduce consist of static, spherically symmetric mappings of this
infinitely large hyperplane plus a point onto the transverse manifold which,
when considering the background in eq.~(\ref{metg2}), is the sphere $\mathbf{S}^{d-p-2}$ 
with radius given by eq.~(\ref{radius}). These solutions are classified according to the homotopy
class
\begin{equation}
\label{map12}
\Pi_{q}(\mathbf{S}^{d-p-2})
\end{equation}
to which they belong.
Let us recall the following identities which we are going to use repeatedly in
the rest of the paper:
\begin{equation}
\label{map1}
\Pi_{q}({\bf S}^{d-p-2})=\left\{ \begin{array}{lll}
0 &{\rm for}\; d-p-2 > q \\
{\bf Z} &{\rm for}\; d-p-2=q.
\end{array} \right.
\end{equation}
We consider implementing a mapping characterized by the homotopy class 
eq.~(\ref{map12}) when $q=d-p-2$. This is accomplished by using an ansatz which associates every point
on the $q$-dimensional hyperplane to point(s) on the compact manifold
$\mathbf{S}^{q}$.\footnote{We consider in detail only simple mappings with 
$q=d-p-2$ since these are the only ones for which we could find a simple ansatz.} 
More specifically, if every point on the hyperplane is associated with $\omega$ points on
the compact transverse space, then the solution is in the class: $\Pi
_{p}(\mathbf{S}^{p})=\omega$. The latter is the winding number,
\textit{i.e.,}\ the number of times the brane wraps the compact manifold.
The metric eq.~(\ref{metg2}) can then be conveniently written in the form
\begin{equation}
ds^{2}=\frac{1}{H(\rho)^{k_{1}}}\left[  -dt^{2}+dr^{2}+r^{2}d\Omega_{q-1}%
^{2} + dx^{I}dx^{I}\right]  +H(\rho)^{k_{2}}\left[  d\rho^{2}+\rho^{2}\left(  d\xi^{2}%
+\sin^{2}\xi d\bar{\Omega}_{q-1}^{2}\right)  \right]  , \label{metg3}%
\end{equation}
where $x^{I}$ ($I=q+1,...,p$) represents directions that are transverse to the world-volume of the probe.
For unit winding number, we use the time-independent spherically symmetric ansatz
\begin{equation}
\Omega_{q-1}=\bar{\Omega}_{q-1}, \label{angles}%
\end{equation}%
\begin{equation}
\xi=F(r), \label{profil}%
\end{equation}
where $F(r)$ will be called the profile angle for reasons that will become
obvious below. It takes values from 0 to $\pi$ while the coordinate $r$ varies
from $+\infty$ to 0 on the world-volume. We supplement the ansatz associated with
eqs.~(\ref{angles}) and (\ref{profil}) with the requirement that the probe be fixed
at $\rho = {\rm const.}$\footnote{This, striclty speaking, can only be valid when the system
preserves some supersymmetries. For a BPS configuration, we take the point of view that the 
topological excitation on the probe, being only a small perturbation, will not 
modify the boundary conditions on the D-brane world-volume so as
to destabilize the system.} 
The resulting configuration is
associated with a conserved topological charge, \textit{i.e.,}\ the winding
number $\omega$. The spherically symmetric ansatz maps the `boundary' of the
$q$-dimensional hyperplane (the region $r\rightarrow\infty$) to a single point
\textit{i.e.}, the north pole of the compact $\mathbf{S}^{q}$ (the point
associated with $\xi=0$). The remaining points on the plane, \textit{i.e.,}\
$\mathbf{R}^{q}-\{\infty\}$, are mapped to points on a compact $q$-sphere minus
its north pole.

In order to understand the physical significance of the solutions associated
with such mappings, let us consider a toy example. Suppose that $F(r)$ is a
monotonically decreasing function of $r$ with boundary values $F(0)=\pi$ and
$F(\infty)=0$. Above some critical radius $r_{c},$ the profile angle may be
considered as zero for all practical means. Accordingly, the map associated
with (\ref{angles}) and (\ref{profil}) then implies that all points for which
$r>r_{c}$ are mapped to a single point on the compact manifold $\mathbf{S}%
^{q}$, \textit{i.e.,}\ the north pole. The points for which $r\leq r_{c}$ are
then in one-to-one correspondence with the points on $\mathbf{S}^{q}$ minus
the north pole. In other words, given the radial quantity $r_{c}$ is small
(for example, it could be of the order of the string length, $r_{c}\sim l_{s}%
$) the map associated with $\Pi_{q}(\mathbf{S}^{q})=1$ corresponds to a
D-brane wrapping a co-cycle $\mathbf{S}^{q}$ but only locally, \textit{i.e.,}%
\ in the region $r\leq r_{c}$. For $r>r_{c}$ the brane probe is point-like
on the transverse $\mathbf{S}^{q}$ and it therefore breaks invariance under
the $SO(q+1)$ group associated with the compact target space, 
a symmetry which is restored in the region $r\leq r_{c}$.

We now consider {\small LWB}'s in a more concrete manner. First, let us evaluate the
gravitational field induced on a D$q$-brane probe in the background geometry
(\ref{metg2}),
\begin{equation}
P[G]_{00}=-\frac{1}{H(\rho)^{k_{1}}},
\end{equation}%
\begin{equation}
P[G]_{rr}=\frac{1}{H(\rho)^{k_{1}}}+\rho^{2}H(\rho)^{k_{2}}F^{\prime}{}^{2},
\end{equation}
where a `prime' denotes a derivative with respect to $r$. The components associated with
the ($q-1$) variables ($\theta_{1}$,...,$\theta_{q-2}$,$\phi$) on the
world-volume of the D-brane are
\begin{equation}
P[G]_{\theta_{1}\theta_{1}}...P[G]_{\theta_{q-2}\theta_{q-2}}P[G]_{\phi\phi
}=\sin^{2q-4}\theta_{1}...\sin^{2}\theta_{q-2}\left(  \frac{r^{2}}%
{H(\rho)^{k_{1}}}+\rho^{2}H(\rho)^{k_{2}}\sin^{2}F\right)  ^{p-1},
\end{equation}
where the angular variables are those defined on a $q$-dimensional plane with $\phi$ an
azimuthal angle. The solution is time-independent so the energy functional is
simply $E=-\int d^{q}\sigma\mathcal{L}_{DBI}$. The equations of motion
associated with the profile angle $F(r)$ are obtained by varying the time
independent action with Dirichlet boundary conditions $F(0)=\pi$ and
$F(\infty)=0$,%
\begin{align}
0  &  =-(q-1)\sin F\cos F-(q-1)\rho^{2}H(\rho)^{k_{1}+k_{2}}F^{\prime}{}%
^{2}\sin F\cos F+(q-1)rF^{\prime}+\nonumber\\
&  (q-1)\rho^{2}H(\rho)^{k_{1}+k_{2}}rF^{\prime}{}^{3}+F^{\prime\prime}\left(
r^{2}+\rho^{2}H(\rho)^{k_{1}+k_{2}}\sin^{2}F\right)  . \label{chiral}%
\end{align}
At this point, it should become obvious that the topological nature of the
solution, the spherically symmetric ansatz for the $\omega=1$ solution and,
finally, this last non-linear equation, are all reminiscent of the Skyrme 
model\cite{skyrme1,skyrme2,adkins-nappi-witten}
where one can also define a similar profile angle $F(r)$. Yet there are
significant differences: The Skyrme model leads to three-dimensional
solitons interpreted as baryons (two-dimensional for the so-called baby-skyrmions)
while our solitons extend in $q$ spatial dimensions. Skyrme introduced an ad-hoc
interaction term in his model whereas the interactions here are gravitational
in nature. The non-polynomial nature of the Lagrangian leads to a
more complex equation for $F(r)$. Moreover, the complexity of eq.
(\ref{chiral}) could be an obstacle to finding topological solitonic
solutions and indeed there is little hope of finding an exact analytical
solution even if the mapping with $\Pi_{q}(\mathbf{S}^{q})=\mathbf{Z}$ implies
the existence of such solutions. We must therefore resort to numerical
techniques to find solutions for different values of $q$. Fortunately, it
turns out that the techniques used to solve the profile angle for the Skyrme
model are appropriate in our case as well \cite{adkins-nappi-witten}. First,
because of the singular behavior of some terms in the equation at $r=0$, one
introduces an analytical solution valid for small $r,$ say $r<r_{0}$, with
the form $F(r)=$ $\pi-ar$ where $a=F^{\prime}(0).$ The rest of the solution is
integrated numerically ajusting the parameter $a$ so that $F(\infty)=0$. This
procedure reveals that the energy is sensitive to the choice of $r_{0}$ and
turns out to be minimum in the limit $r_{0}\rightarrow 0$, in which case the
contribution comes entirely from the region $r<r_{0}.$ This corresponds to a
point-like configuration or an exact profile angle which is the step function
\FIGURE{\epsfig{file=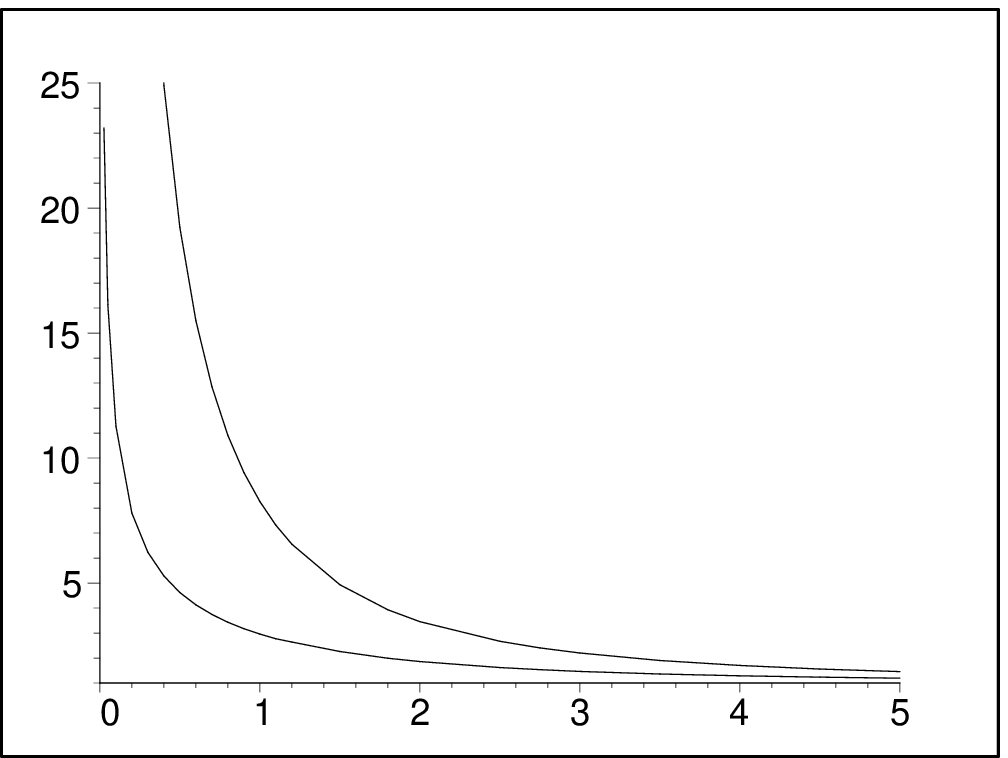, width = 6cm}
\caption{The ratio $E/V_{{\bf S}^{p}%
}(\rho)$ of the configuration for $T_{p}=1$ and 
$H(\rho)=1$ as a function of $R$. 
The lower (upper) curve 
corresponds to the case $p=3$ ($p=4$).
}%
\label{g1}} 
\begin{equation}
F(r)\sim\frac{\pi}{2}\left(  1+\theta(r)\right)  \;\;\;\;\;\;r\geq0,
\end{equation}
where $\theta(x)=1$ for $x\leq0$ and $\theta(x)=-1$ for $x>0$. Consequently,
the minimal energy solution corresponds to a homotopy mapping of the points in
the $q$-dimensional plane onto the north pole of the compact sphere
$\mathbf{S}^{q}$ and of the origin of ${\bf R}^{p}$ onto the other points
of the target sphere. It would be wrong to conclude that this configuration
corresponds to no wrapping at all. In fact, the energy of the configuration
is
\begin{equation}
E=T_{q}H(\rho)^{k_{3}-\frac{pk_{1}}{2}}V_{{\bf S}^{q}}(R), \label{genenergy}%
\end{equation}
where the volume $V_{{\bf S}^{q}}(R)$ is calculated using the radius $R$ defined in eq.~(\ref{radius}).
One should regard this configuration as a local wrapping corresponding to
taking the continuous limit $r_{c}\rightarrow 0$ in the toy example
described earlier. The finiteness of the energy in this limit is attributed to
the fact that it is impossible to change the topological sector of an
excitation by any continuous process. In the next section we present evidence
that the $\alpha^{\prime}$ curvature corrections to the {\small DBI} action might allow the
wrapping to expand to a size $r_{c}\sim\sqrt{\alpha^{\prime}}=l_{s}$.

\section{Evidence for delocalization of the wrapping}
\label{evidence} The {\small DBI} action with the U(1) gauge field turned off is the
first term of an expansion in the parameter $\alpha^{\prime}$. 
In other words, the {\small DBI} action in
eq.~(\ref{bi2}) constitutes the $\alpha^{\prime}=0$ limit of that expansion,
\ie the field theory limit. This raises the question of how {\small LWB}'s are affected
when $\alpha^{\prime}$ corrections are switched on. In a sense, it is natural
that the $\alpha^{\prime}=0$ limit leads to a singular phenomena since the excitations
are then derived from a theory where the fundamental length scale has been set to zero.
Saying that it is necessary to
include the $\alpha^{\prime}$ corrections is a statement that the curvatures
involved in the problem (in our case the curvature of the transverse sphere), 
are large compared to $1/\alpha^{\prime}$.
For now, we take the point of
view that the curvature is large but still small enough that only one extra
term needs be considered in the $\alpha^{\prime}$ expansion. This assumption,
and other aspects of the curvature corrections, are discussed
in section \ref{discussion}, but for now is motivated by a
rather practical reason: only one term in the expansion has been calculated, to
our knowledge. In bosonic string theory, the $\mathcal{O}(\alpha^{\prime})$
correction has been found to be \cite{corley} (for totally geodesic branes),
\begin{equation}
-\alpha^{\prime}T_{q}\int d^{q+1}\sigma\sqrt{-|P[G]_{ab}|}%
{\cal R},\label{correction1}%
\end{equation}
where ${\cal R}$ is the Ricci scalar evaluated with the induced metric
$P[G]_{ab}$. For the D-branes in superstring theory, the $\mathcal{O}%
(\alpha^{\prime})$ correction vanishes identically but the $\mathcal{O}%
(\alpha^{\prime}{}^{2})$ correction is a sum of terms involving
quadratic products of the Riemann and Ricci tensors as well as the Ricci
scalar \cite{bachas,fotopoulos,wyllard1,wyllard2}.
\FIGURE{\epsfig{file=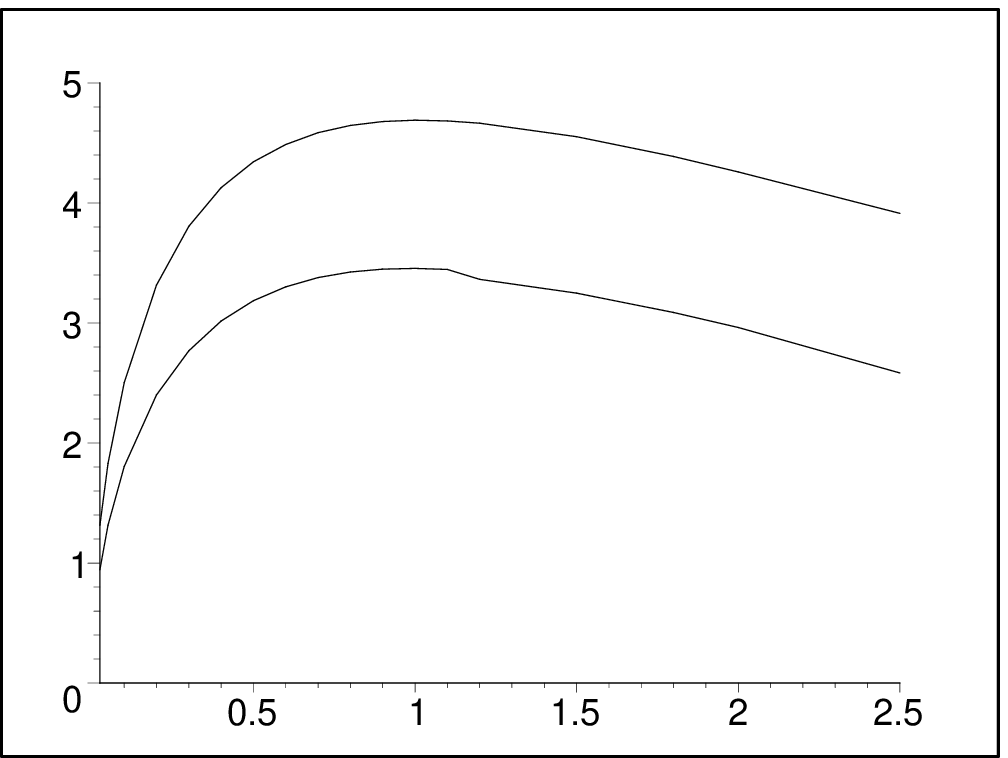, width = 5cm}%
\caption{The size $r_{s}$ of the configuration as a function of $\rho$  
for  $p=3$ (lower curve) and $p=4$.  As $\rho$ gets larger the 
size decreases to a point-like object. On the other hand, the size of the configuration 
reaches a maximum near $\rho=1$ in both cases.
}%
\label{g4}} 

As an example, we consider adding the term (\ref{correction1}) to the {\small DBI}
action eq.~(\ref{bi2})\footnote{Our attempts to include the ${\cal O}(\alpha^{\prime}{}^{2})$ as well
have failed due to the highly complicated nature of the corresponding differential equation.}. 
Once again, we use the spherically symmetric ansatz
corresponding to (\ref{angles}) and (\ref{profil}) and minimize the resulting
energy functional which leads to a differential equation for the profile
$F(r).$ The new terms that need to be added to eq.~(\ref{chiral}), although easily derived, are
numerous and complicated so we do not write them
here. Integrating the differential equation using the procedure described
above, it becomes clear that the profile is not localized at a point anymore
but rather is a monotonically decreasing function of $r$. In other words, a
delocalization of the wrapping has been induced by the $\mathcal{O}%
(\alpha^{\prime})$ correction. In Fig.~\ref{g1}, we illustrate how the energy
of the configuration, shown here in terms of the ratio $E/V_{{\bf S}^{q}}(R),$
depends on $\rho$ for a simple non-physical case where $T_{q}=1$ and
$H(\rho)\equiv1$ so that the rescaled radius $R=\rho$ here. The lower (upper)
curve corresponds to the case $p=3$ ($p=4$). In both cases we find that for
large values of the target space volume (which corresponds to the 
$\mathcal{O}(\alpha^{\prime})$ correction becoming smaller and
smaller), the ratio $E/V_{{\bf S}^{p}}(R)$ approaches 1 which corresponds to the
result in eq.~(\ref{genenergy}). On the other hand, a close analysis reveals
that the energy $E$ tends to zero as $\rho$ reaches zero. Delocalization of
the configurations becomes evident when one examines how the energy is
distributed in space. For that purpose we introduce the ``size'' of the
configuration to be defined, somewhat arbitrarily, as the radius $r_{s}$ for which the
energy density reaches 1\% of its maximum value. The results are presented in
Fig.~\ref{g4} where the mapping becomes increasingly local for large values
of the target space volume. However, the size of the configuration $r_{s}$
reaches a maximum near $\rho=1$ in both cases. Finally, for small values of
the target space volume where the $\mathcal{O}(\alpha^{\prime})$ corrections
are increasingly large the configuration tends to be point-like again. Of
course this last observation is only valid when higher order corrections in
$\alpha^{\prime}$ are not taken into account and clearly these are not negligible
in a regime where $V_{{\bf S}^{q}}$ is small.

Although we have gone through the process of solving numerically the profile
differential equation, one may, in certain cases, look at the problem using a more intuitive
approach. The reader can find in the appendix an example illustrating the
delocalization effect by way of using a specific ansatz for the profile $F(r)$
and variational approach. The results are of course not exact but a similar
behavior to what we have found numerically is observed for $p=3$.

\section{Applications in string and M-theory}
\label{applications}

We present examples in string and M-theory where the formalism introduced in
section \ref{description} is applicable. We mostly consider brane probes
interacting with fields that are sourced by supergravity solutions
descending from either string or M-theory. Firstly, we present Type
II$\,$a,b superstring theory examples characterized by non-trivial homotopy
mappings of the form $\Pi_{q}(\mathbf{S}^{p})$. We then consider
applications of the {\small LWB} formalism in M-theory, in critical bosonic string theory 
and, finally, in the speculative 
bosonic M-theory proposed in ref.~\cite{horowitz}. The examples fall in two
categories: (1) those with $q=p$ to which the mathematical analysis of section 
\ref{description} applies directly and (2) those for which $q\neq p$ but that are nevertheless
associated with a non-trivial homotopy mapping. 

\subsection{Superstring theory examples}
\label{stringex} 

The low energy limit of the type II$\,$a,b superstring theories
are the ten-dimensional type II$\,$a,b supergravity theories. The type II$\,$a
theory contains D$p$-branes with $p$ even while the type II$\,$b theory contains branes
associated with $p$ odd. The supergravity (low energy) manifestation of these
objects, the $p$-branes, are characterized by an $SO(1,p)\times SO(9-p)$
symmetry which is manifest from the gravitational field they
generate\cite{strominger}\footnote{The notation used is that of
ref.~\cite{peet}.},
\begin{equation}
ds^{2}=\frac{1}{H_{p}(\rho)^{1/2}}\left[  -dt^{2}+dx^{m}dx^{m}\right]
+H_{p}(\rho)^{1/2}\left[  d\rho^{2}+\rho^{2}d\Omega_{8-p}^{2}\right]  ,
\label{pmetric}%
\end{equation}%
\begin{equation}
e^{\Phi}=H_{p}(\rho)^{\frac{1}{4}(3-p)}. \label{pdilaton}%
\end{equation}
These fields correspond to the general solution associated with eqs.~(\ref{metg}) and (\ref{dilg})
where $k_{1}=k_{2}=1/2$ and $k_{3}=(p-3)/4$. 
Like their microscopic realization, the $p$-branes are natural
sources for the R-R form field\cite{polchinski3}
\begin{equation}
C_{01...p}(\rho)=\frac{1}{g_{s}}\left[  1-\frac{1}{H_{p}(\rho)}\right].
\label{prr}%
\end{equation}
The corresponding Chern-Simons term does not affect the physics of a {\small LWB} which involves couplings
of background fields to the world-volume scalars $X^{i}$.\footnote{There are two instances where
the Chern-Simons couplings might be of relevence: (1) when the R-R field depends on the world-volume
coordinates. Then, a Taylor expansion in the Chern-Simons action will result in new couplings with
the world-volume scalars, (2) in the non-Abelian case, 
couplings of the matrix-valued scalars might be induced by the presence of a higher form field\cite{myers1}.} 
$H_{p}(\rho)$ is a harmonic function in the space transverse to
the world-volume of the $p$-brane,
\begin{equation}
H_{p}(\rho)=1+\frac{c_{p}g_{s}Nl_{s}^{7-p}}{\rho^{7-p}},
\end{equation}
where $c_{p}$ is a dimensionless constant\cite{peet}, $g_{s}$ is the string
coupling and $N$ is the flux of the R-R field (\ref{prr}) through the
$\mathbf{S}^{8-p}$ factor of the metric. Strictly speaking, the field expressions (\ref{pmetric}),
(\ref{pdilaton}) and (\ref{prr}) are valid only for $p<7$. As it stands, we
will not need the expressions corresponding to $p\geq7$ in what follows.

Our approach consists in probing the $p$-brane geometries with fundamental D$q$-branes in
such a way as to determine for which values of $q$ a {\small LWB} excitation can exist. There are
three requirements:
\begin{enumerate}
\item $q\leq p$: because only then can the world-volume of the probe
be parallel to the world-volume of the $p$-brane. It is crucial that the D-brane
be orthogonal to the $\mathbf{S}^{8-p}$ factor.
\item $q\geq8-p$: otherwise it is guarenteed that the mapping $\Pi_{q}({\bf S}^{8-p})$ is trivial.
\item If $p$ is odd (even) then $q$ must be odd (even) since D-branes
of opposite parity cannot exist in the same theory.
\end{enumerate}
Leaving aside for now issues of stability,
there are ten configurations, in either type II$\,$a or type II$\,$b
string theory, which satisfy these three criteria. To enumerate those cases, we
introduce the notation D$q$/$p$\;($\Pi_{q}(\mathbf{S}^{8-p})$) which refers to a
system composed of a D$q$-brane probe in the supergravity background generated
by a $p$-brane. The corresponding homotopy mapping $\Pi_{q}(\mathbf{S}^{8-p})$
is assumed to be non-trivial. The potential superstring theory examples are%
\[%
\begin{tabular}
[c]{cc}%
Type II$\,$a & Type II$\,$b\\\hline
D2/6($\Pi_{2}(\mathbf{S}^{2})$), D4/4($\Pi_{4}(\mathbf{S}^{4})$), &
D1/7($\Pi_{1}(\mathbf{S}^{1})$), D3/5($\Pi_{3}(\mathbf{S}^{3})$),\\
D4/6($\Pi_{4}(\mathbf{S}^{2})$), D6/6($\Pi_{6}(\mathbf{S}^{2})$), &
D3/7($\Pi_{3}(\mathbf{S}^{1})$), D5/5($\Pi_{5}(\mathbf{S}^{3})$),\\
& D5/7($\Pi_{5}(\mathbf{S}^{1})$), D7/7($\Pi_{7}(\mathbf{S}^{1})$).
\end{tabular}
\]

Recall that the setup we are considering is a D$q$-brane probe placed at a fixed distance from
a $p$-brane. Before considering {\small LWB} excitations on the probe, it would
seem appropriate to ask whether if the system is initially (before the excitation
is made to appear) stable or not. 
Of course, not all of the systems mentionned in the list above are stable. 
This can be verified straightforwardly
by calculating whether the force (due to graviton,
dilaton and R-R field exchange) exerted by the $p$-brane on the probe vanishes
or not. It turns out that co-dimension 0 (mod 4) systems are stable (BPS\footnote{More precisely, 
BPS means that the brane configuration preserves 1/2
of the bulk supersymmetry. The stability of the system can be traced to that fact.
Conversely, the unstable systems we are refering to do not preserve any 
supersymmetries.}),
co-dimension 2 systems are attractive and co-dimension 6 objects are
repulsive. Consequently, the only potentially stable systems in the list are
\begin{align}
&  \text{D2/6}(\Pi_{2}(\mathbf{S}^{2})),\;\;\;\text{D4/4}(\Pi_{4}%
(\mathbf{S}^{4})),\;\;\;\text{D6/6}(\Pi_{6}(\mathbf{S}^{2})),\\
&  \text{D3/7}(\Pi_{3}(\mathbf{S}^{1})),\;\;\;\text{D5/5}(\Pi_{5}%
(\mathbf{S}^{3})),\;\;\;\text{D7/7}(\Pi_{7}(\mathbf{S}^{1})).\nonumber
\end{align}
The other systems are either attractive, causing the probe to collide with the
$p$-brane, or repulsive. 
In section \ref{description} we have described a
technique allowing one to calculate the energy of the configurations for which
$q=8-p$. Among the stable configurations, those satisfying this equality are the following type II$\,$a
configurations,
\begin{eqnarray}
\text{D2/6}(\Pi_{2}(\mathbf{S}^{2})) \;\; {\rm with}\;\; E = T_{2} H_{6}(R)^{-\frac{3}{4}}V_{{\bf S}^{2}}(R), \\
\text{D4/4}(\Pi_{4}(\mathbf{S}^{4}))\;\; {\rm with}\;\; E = T_{4} H_{4}(R)^{-\frac{3}{4}}V_{{\bf S}^{4}}(R),
\end{eqnarray}
where $R=\rho H_{p}^{\frac{1}{4}}$ with $p=6$ and $p=4$ respectively.
There are also configurations among
the unstable ones which are such that $q=8-p$. Those correspond to the type II$\,$b
systems
\begin{eqnarray}
\text{D1/7}(\Pi_{1}(\mathbf{S}^{1}))\;\; {\rm with}\;\;  E = T_{1} H_{7}(R)^{-\frac{3}{4}}V_{{\bf S}^{1}}(R), \\
\text{D3/5}(\Pi_{3}(\mathbf{S}^{3}))\;\; {\rm with}\;\;  E = T_{3} H_{5}(R)^{-\frac{3}{4}}V_{{\bf S}^{3}}(R),
\end{eqnarray}
which are respectively repulsive and attractive. One can verify that the energy the {\small LWB} excitations
considered is always a monotonically increasing function of $\rho$. 

\subsection{Example involving a M-brane probe}

The fundamental branes of M-theory are M2- and M5-branes. The low energy
action of M-theory corresponds to eleven-dimensional supergravity with a
spectrum comprising of the graviton and a fundamental three-form field. The
M2-brane is electrically charged with respect to the three-form field while
the M5-brane is a magnetic source for it. The geometry generated by a
distribution of $N$ parallel M2-branes is associated with the
metric,
\begin{equation}
ds_{11}^{2}=\frac{1}{H_{M2}(\rho)^{2/3}}\left[  -dt^{2}+dx^{i}dx^{i}\right]
+H_{M2}(\rho)^{1/3}\left[  d\rho^{2}+\rho^{2}d\Omega_{7}^{2}\right]  ,
\end{equation}
which has a manifest $SO(1,2)\times SO(8)$ symmetry. The harmonic function is
\begin{equation}
H_{M2}(\rho)=1+\frac{\rho_{-}^{6}}{\rho^{6}},
\end{equation}
where $\rho_{-}^{6}\sim N$. We consider probing this `macroscopic' geometry
with fundamental M2- and M5-branes. We are interested in finding whether a {\small LWB}
can be excited on the probes. Having a locally wrapped M5-brane probe is excluded as it does
not even satisfy the criteria (1) introduced in section \ref{stringex}. Locally
wrapping a M2-brane probe is excluded as well but in this case because such a
configuration would violate criteria (2) since $\Pi_{2}(S^{7})$ is trivial.

We now consider the metric associated with a distribution of $N$
parallel M5-branes,
\begin{equation}
ds_{11}^{2}=\frac{1}{H_{M5}(\rho)^{1/3}}\left[  -dt^{2}+dx^{i}dx^{i}\right]
+H_{M5}(\rho)^{2/3}\left[  d\rho^{2}+\rho^{2}d\Omega_{4}^{2}\right]  ,
\label{m5metric}%
\end{equation}
which has a manifest $SO(1,5)\times SO(5)$ symmetry. The harmonic function is
\begin{equation}
H_{M5}(\rho)=1+\frac{\rho_{-}^{3}}{\rho^{3}}.
\end{equation}
In this case, a M2-brane probe cannot be locally wrapped because it violates
criterion (2) since $\Pi_{2}(S^{4})$ is again trivial. A M5-brane probe in the background
geometry (\ref{m5metric}) can be locally wrapped since $\Pi_{5}(S^{4})$ is non-trivial.
In the absence of a convenient ansatz to study the dynamical behavior of the
resulting configuration no conclusion can be drawn, 
though there is no reason to believe the system would behave differently
than the type II$\,$a D4/4 configuration for example. The M5/5 system
has the interesting feature of being BPS. Note that
if a delocalization process occurs due to higher order curvature corrections,
it is in this case expected to be of the order of the eleven-dimensional Planck length, $l_{P}$,
which is the fundamental scale in the low energy limit of M-theory.

\subsection{Locally wrapped bosonic branes}

We now comment on the possibility of occurence of {\small LWB} excitations in
critical bosonic string theory and in the much more speculative bosonic M
theory proposed in ref.~\cite{horowitz}. 

\subsubsection{D-branes in critical bosonic string theory}

The D-branes of bosonic string theory are plagued with an open string tachyon
making them unstable\cite{GSW,polchinski2}. While this fact may be a blessing in
disguise\cite{sen1,sen2}, it is not pleasant for our immediate purposes. In
fact, the setup we would like to study is a bosonic D$q$-brane probing placed in a
background containing a geometrical factor ${\bf S}^{p}$ in such a way that $\Pi
_{q}(\mathbf{S}^{p})$ is non-trivial\footnote{Of course, one would ideally require such a
background to be generated by a configuration constructed out of objects coming from
bosonic string theory. We will be content here in probing an arbitrary yet
stable background.}. The D-brane probe action will include couplings of the
open string fields with the world-volume closed string tachyon. This
clearly leads to an instability unless the tachyon is assumed to have been
taken care of by some unknown mechanism. There is also the problem of the
closed string tachyon which is, in a sense, more
pressing as it appears to make the whole theory, not just the D-branes, unstable.

Ignoring the tachyon problems, one might imagine inserting a D$q$-brane probe
in, for example, the background AdS$_{p+2}\times {\bf S}^{p}\times {\bf S}^{24-2p}$ with
$d=p+q=26$. These spacetimes have been shown to be stable under small field
perturbations (without any supersymmetry) for $q>8$\cite{bosback1,bosback2}.
The probe action in this case has the same form as the one used in superstring
theory, \textit{i.e.,}\ eq.~(\ref{bi}) but this time there is no Chern-Simons
term. We assume that the world-volume of the probe is parallel to $q+1$
directions of the AdS factor. Consequently, the brane will be allowed to be locally
wrapped if
\begin{itemize}
\item $q\leq p$,
\item $\Pi_{q}({\bf S}^{p})\neq0$ and/or $\Pi_{q}({\bf S}^{24-p})\neq0$.
\end{itemize}
There is, {\it a priori}, no reason for the probe not to wrap both spherical co-cycles if the
two homotopy mappings are non-trivial. We have given evidence in section \ref{evidence}
that the {\small LWB} excitations in bosonic string theory might become delocalized when the $\alpha^{\prime}$
curvature corrections to the probe action are included.

\subsubsection{Branes in bosonic M-theory}

We now present an example of a {\small LWB} in the context of the speculative
27-dimensional bosonic M-theory proposed by Horowitz and
Susskind\cite{horowitz}\footnote{See however ref.~\cite{west} where it is
argued, based on a coset symmetry analysis, that if a 27-dimensional bosonic M
theory exists it does not have a simple local low energy effective action.}.
They proposed that closed bosonic string theory is actually a compactification
of a 27-dimensional theory on the orbifold $S^{1}/\mathbf{Z}_{2}$ with no extra
degrees of freedom living at the fixed points. The authors of
ref.~\cite{horowitz} have argued that the open string tachyon instability may
be removed in the strong coupling limit, allowing them to study branes in
this theory. The proposed low energy action for bosonic M-theory is
\begin{equation}
S = \int d^{27}x \sqrt{-g} \left[  R - \frac{1}{48} F_{\mu\nu\rho\lambda}
F^{\mu\nu\rho\lambda} \right]  ,
\end{equation}
where $F=dC$. The fundamental fields are the graviton and the supergravity
three-form which are sourced by M2- and M21-branes. These objects are the
extremal limits of black brane solutions\cite{gibbons}. The fields associated
with the electrically charged bosonic M2-brane are
\begin{equation}
\label{bosM1}ds^{2} = \frac{1}{H_{M2}(\rho)^{2/3}} \left[  -dt^{2} +
dx^{m}dx^{m} \right]  + H_{M2}(\rho)^{1/11} \left[  d\rho^{2} + \rho^{2}
d\Omega_{23} \right]  ,
\end{equation}
where
\begin{equation}
H_{M2}(\rho) = 1 + \left(  \frac{\rho_{-}}{\rho}\right)  ^{22},
\end{equation}
and $\rho_{-}^{22}\sim N$. The four-form under which the M2-brane is
electrically charged is
\begin{equation}
^{*}F = N l_{P}^{22} \epsilon_{23},
\end{equation}
where $N$ is the number of fundamental branes forming the `macroscopic'
configuration, $l_{P}$ is the 27-dimensional Planck length, and $\epsilon_{23}$ is
the volume form on a unit $\mathbf{S}^{23}$. The metric
associated with a distribution of $N$ M21-branes is
\begin{equation}
\label{bosM2}ds^{2} = \frac{1}{H_{M21}(\rho)^{1/11}} \left[  -dt^{2} +
dx^{i}dx^{i} \right]  + H_{M21}(\rho)^{2/3} \left[  d\rho^{2} + \rho^{2}
d\Omega_{4} \right]  ,
\end{equation}
where
\begin{equation}
H_{M21}(\rho) = 1 + \left(  \frac{\rho_{-}}{\rho}\right)  ^{3},
\end{equation}
and $\rho_{-}^{3}\sim N$. The four-form under which the M5-brane is
magnetically charged is
\begin{equation}
^{*}F = N l_{P}^{3} \epsilon_{4}.
\end{equation}
The extremal branes are completely stable quantum mechanically and there is no
force between pairs of extremal branes of the same dimensionality in complete
analogy with the branes in superstring and M-theory.

We now consider whether M2- and M21-brane probes can be locally wrapped when
placed in the backgrounds corresponding to eqs.~(\ref{bosM1}) and (\ref{bosM2}). Following the analysis
of the previous sections, we see that only the M21-brane in
the background of a 21-brane supergravity solution can be locally wrapped. In
fact, the M21-brane can wrap the $\mathbf{S}^{4}$ since $\Pi_{21}%
(\mathbf{S}^{4})$ is a non-trivial homotopy mapping. Moreover, the system is stable because the attraction
due to graviton exchange is exactly cancelled by the repulsion induced by the
exchange of fundamental three-form field quanta. This example is in complete analogy with
the M5/5 configuration we found in M-theory.

\section{Discussion}
\label{discussion}
In this work we have considered D- and M-branes locally wrapping factors
with spherical symmetry in geometries that are generated by distributions of other branes.
The examples we solved for in detail involve simple homotopy
mappings of the form $\Pi_{q}(\mathbf{S}^{q})=\mathbf{Z}$. The main reason for
treating mathematically only these cases is that we know of a simple way to write the associated
spherically symmetric ansatz for the mapping of the $q$-dimensional spatial
part of the probe world-volume onto the compact spherical target manifold. There
obviously exist other spacetimes of interest to string and M-theory where
brane probes could locally wrap more complicated compact manifolds that we here denote
$\mathbf{X}_{p}$. The main criterion for that to be possible is that the homotopy mapping
\begin{equation}
\Pi_{q}(\mathbf{X}_{p})
\end{equation}
be non-trivial.
To convince oneself that {\small LWB} excitations might be quite common, we refer the
reader to ref.~\cite{nak} where homotopy mappings to target spaces such as
$E_{6}$, $E_{8}$ or $G_{2}$ are considered. 

The excitation we described is a D$q$- or M$q$-brane that
is wrapping the compact manifold $\mathbf{X}_{p}$ at one point of its
world-volume. More concretely, the resulting configuration is composed of a
fundamental D$q$- or M$q$-brane out of which another brane with $p$ spatial
dimensions protrudes at right angles. We also gave evidence that the wrapping
might not be local if one is willing to take into account
higher order curvature corrections to the brane probe action. It is
conceivable that these will lead to a delocalization of the wrapping on a
length scale which of the order of the fundamental scale of the theory,
\textit{i.e.,}\ $\sqrt{\alpha^{\prime}}=l_{s}$ or $l_{P}$, depending on whether one
is considering branes in superstring or M-theory.

The {\small DBI} action is a generalized ($q+1$)-dimensional volume which
transpires in the expression for the energy of a {\small LWB} excitation
eq.~(\ref{genenergy}). Given the analysis performed for local wrappings based
on the homotopy mapping $\Pi_{q}(\mathbf{S}^{q})$, we can guess the generic
form of the energy of an excitation associated with an arbitrary non-trivial mapping $\Pi
_{q}(\mathbf{{\bf X}}_{p})$:
\begin{equation}
E=(\text{red shift})T_{q}V_{\mathbf{X}_{p}},\label{EV}%
\end{equation}
where $T_{q}$ is the tension of the probe, $V_{\mathbf{X}_{p}}$ is the volume
of the wrapped manifold calculated with a rescaled radius (see,
\textit{e.g.,}\ eq.~(\ref{radius})), and the multiplicative red shift factor depends on
the nature of the gravitational background probed. 

In the supergravity examples considered, we have assumed that the brane probe was placed at 
a fixed radial distance ($\rho={\rm const.}$) away from the horizon. We then used 
a spherically symmetric ansatz to find a profile function $F(r)$ that minimizes the energy of
the configuration. It is quite possible that this procedure imposes
too much constraint on the resulting brane system. In fact, it would seem natural to expect that the region
where the brane wraps the compact manifold would move in the $\rho$ direction so as to minimize its energy
further. 
We therefore expect that there will be more structure to a {\small LWB}.
This could be checked by relaxing the constraint $\partial_{a}X^{p+1}=\partial_{a}\rho=0$ imposed in
section \ref{description}. Doing so can accomplished by introducing the function
\begin{equation}
h(r) = \rho(r) - \rho_{0},
\end{equation} 
where $\rho_{0}={\rm const.}$ is the initial distance of the probe away from the horizon at $\rho=0$.
Then, one needs not only to minimize the energy of a {\small LWB} with respect to $F(r)$ but
with respect to $h(r)$ with the boundary conditions $h(r\rightarrow +\infty)=0$ as well. Recall that we have found
the energy of a {\small LWB} with $h(r)=0$ to be minimized when the probe is placed as close as possible to the horizon.
We therefore expect that the energy of the brane excitation will be minimized for a negative $h(r)$.
In other words, the resulting configuration will be a brane that locally wraps a compact manifold but
is slightly deformed in the $\rho$ direction toward the horizon. 

The energy
relation eq.~(\ref{EV}) will also get modified when one takes into account curvature
corrections to the much studied {\small DBI} action. Recall that the full D-brane action (with
$B_{ab}=0$, $F_{ab}=0$) can be seen as an expansion in
$\alpha^{\prime}$, the first term being the {\small DBI} action. For
example, we find that including the first order $\alpha^{\prime}$ correction
derived from bosonic string theory implies the following:
\begin{itemize}
\item The modification to the energy of the {\small LWB} is proportional to the
curvature of the target space, \textit{i.e.,}\ the associated Ricci scalar.
Consequently, having transverse manifolds with a large curvature compared to
$1/\alpha^{\prime}$ corresponds to a regime where the $\alpha^{\prime}$ correction may
not be negligible with respect to the $\mathcal{O}(\alpha^{\prime}{}^{2})$
corrections.
\item If the system is in a regime where the curvature of the transverse
space is small in such a way that the $\mathcal{O}(\alpha^{\prime})$
correction is the dominant one, the {\small LWB} excitation is found to expand or
delocalize itself to a size which is of the order of the string length.
\end{itemize}
Consequently, the evidence we gave for a delocalization of the wrapping is not convincing
when the curvature of the target space is large, but given it is small enough
we find that there is definitely a regime for which the $\mathcal{O}%
(\alpha^{\prime})$ correction dominates. For example, consider for simplicity the case 
where $H(\rho)=1$ in eq.~(\ref{metg2}) with a spherical target space. Then, the
curvature is proportional to $1/\rho^{2}$ where $\rho$ is the radius of
$\mathbf{S}^{p}$. The expansion of the D-brane action then
schematically takes the form
\begin{equation}
S_{D-brane}\sim1+a_{1}\frac{\alpha^{\prime}}{\rho^{2}}+a_{2}\frac{\alpha^{\prime}%
{}^{2}}{\rho^{4}}+\;\;...\;,
\end{equation}
where the $a_{i}$'s are dimensionless coefficients determined by string theory
calculations. It is clear, given that the coefficients $a_{i}$ are all of the same
order of magnitude, that the validity of our conclusions regarding the
delocalization of the soliton states is determined by whether $\rho$ is large or not.
For the D-branes in superstring theory, $a_{1}=0$ and the first curvature correction is
the $\mathcal{O}(\alpha^{\prime2})$ one. We have been unsuccessful at
generalizing our numerical calculation to include this correction. The
conclusion we can reach is that it is conceivable that {\small LWB} excitations in
bosonic string theory will expand because of the curvature corrections
although we have no way to predict whether taking into consideration higher
order corrections will ultimately ruin this behavior. We can only take
this bosonic calculation as an indication that a similar phenomenon might
happen for the D-branes in superstring theory and the M-branes in M-theory.

\acknowledgments             The authors would like to thank John Brodie, Cliff Burgess,
Neil Constable, Clifford Johnson, Juan Maldacena, Don Marolf, Konstantin Savvidy 
and David Winters for interesting
conversations. We are especially indebted to 
Rob Myers and Patrick Labelle for their useful input to the project 
during its realization. The authors were
supported in part by NSERC of Canada and Fonds FCAR du Qu\'ebec. FL would like
to express his thanks to the Perimeter Institute as well as to the Departments
of Physics at the Universit\'e Laval and the University of Waterloo for their
hospitality during certain stages of this work.

\appendix            

\section{Atiyah-Manton solitons}
In this appendix we present a scheme which allows one to obtain approximative
solutions to the equations of motion associated with locally wrapped
excitations and yet to easily deduce most of their relevant physical
properties without resorting to a full numerical calculation. Recall that the
contribution of the `non-accelerating' part of the {\small DBI} action to the energy of
a locally wrapped D$p$-brane is\footnote{For simplicity we consider here a 
non-physical background corresponding to setting $H(\rho)=1$ in eq.~(\ref{metg}) 
and with the dilaton $\Phi$ vanishing everywhere.} 
\begin{eqnarray}
E=T_{p}\int d^{p}\sigma \sqrt{g_{\Omega}} \left[  (1+\rho^{2}F^{\prime}{}^{2})(r^{2}+\rho^{2}%
\sin^{2}F)^{p-1}\right]  ^{1/2} - \nonumber \\ T_{p} \int d^{p}\sigma \sqrt{g_{\Omega}} r^{p-1},\label{amendens}%
\end{eqnarray}
where we have substracted the energy on an infinitely extended brane. We have introduced the notation
$\sqrt{g_{\Omega}}$ to denote the volume factor associated with the angular variables on the
world-volume.
\FIGURE{\epsfig{file=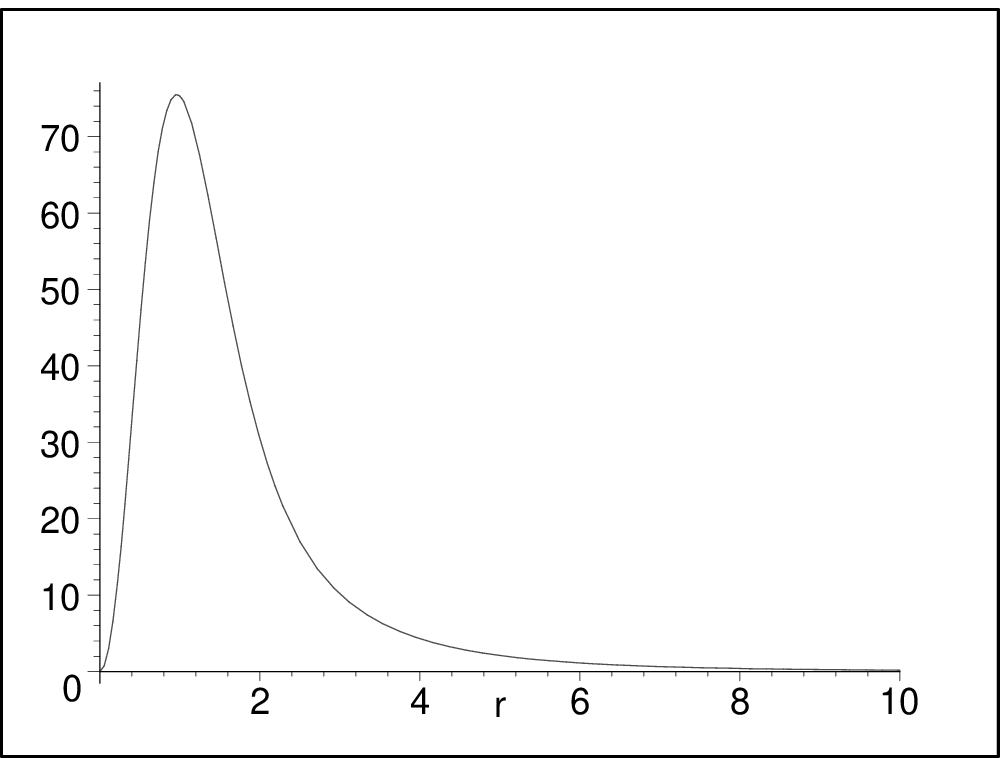, width = 4cm}%
\caption{Energy density
of an unstable locally wrapped D3-brane.
The excitation has a typical world-volume size $L=2\,l_{s}%
$ and the radius of the
spherical co-cyle is $\rho=4\,l_{s}$.}\label{lws1}}
It was proposed in ref.~\cite{dobado} that one can use the Atiyah-Manton ansatz for
$F(r)$ in order to solve the differential equation associated with this
profile. Let us first recall that the Atiyah-Manton ansatz originates from the
computation of the holonomy of instanton solutions and as such, it should
represent a three dimensional object. It is worth pointing out at
this point that while, in principle, one could probably use the ansatz for other
purposes, this is only justified for the $p=3$ case. Applied to the
excitations of interest, the Atiyah-Manton ansatz takes the form
\begin{equation}
F(r)=\omega\pi\left(  1-\frac{1}{\sqrt{1+\frac{L^{2}}{r^{2}}}}\right)  ,
\end{equation}
where $\omega$ is the winding number (associated with the topological charge) and
$L$ characterizes the size of the excitation, as will be made clear in a short
while. We use this ansatz to reproduce the results of ref.~\cite{dobado}, but with
emphasis on the physical properties of the resulting soliton. We will later
introduce a typical $\alpha^{\prime}$ curvature correction and study its effects. Figure \ref{lws1} shows the
energy density as a function of the radial distance from the center of an
unstable excitation for which the typical size is $L=2\,l_{s}$. The radius of
the spherical co-cycle on which the brane is wrapped was arbitrarily chosen to be
$\rho=4\,l_{s}$.
In order to see why this configuration is unstable one needs to vary the
energy functional with respect to the parameter $L$. As it turns out, the
minimum of $E$ is for $L=0$ for all values
of $\rho$. As explained earlier, one should not interpret the $L=0$ states as
being non-solitonic. One should see the $L\rightarrow0$ as a limiting
process in which one stays in the same topological sector (in this case
$\Pi_{3}(S^{3})=1$).

An interesting idea, proposed in ref.~\cite{dobado} in the context of brane-world
scenarios, consists in investigating ways to stabilize the configuration to
$L\neq 0$. We therefore extend the calculation by including the first order
$\alpha^{\prime}$ correction to the {\small DBI} action obtained from critical bosonic string
theory \cite{corley}. The energy associated with this new term is
found to be
\begin{equation}
E_{\alpha^{\prime}}=\alpha^{\prime}T_{p}(p-1)\int d^{p}\sigma \sqrt{g_{\Omega}}
\left[  (1+\rho^{2}F^{\prime}{}^{2})(r^{2}+\rho^{2}%
\sin^{2}F)^{p-1}\right]  ^{1/2} 
\frac{M G^{\prime
}+(p-2)M^{\prime}G(G-1)}{M^{2}M^{\prime}G^{2}},\label{corrb}%
\end{equation}
where we have introduced, for simplicity,
\begin{equation}
M^{2}=r^{2}+\rho^{2}\sin^{2}F,\;\;\;\;\;G=\frac{1+\rho^{2}(F^{\prime
})^{2}}{M^{\prime}{}^{2}}.
\end{equation}
\FIGURE{\epsfig{file=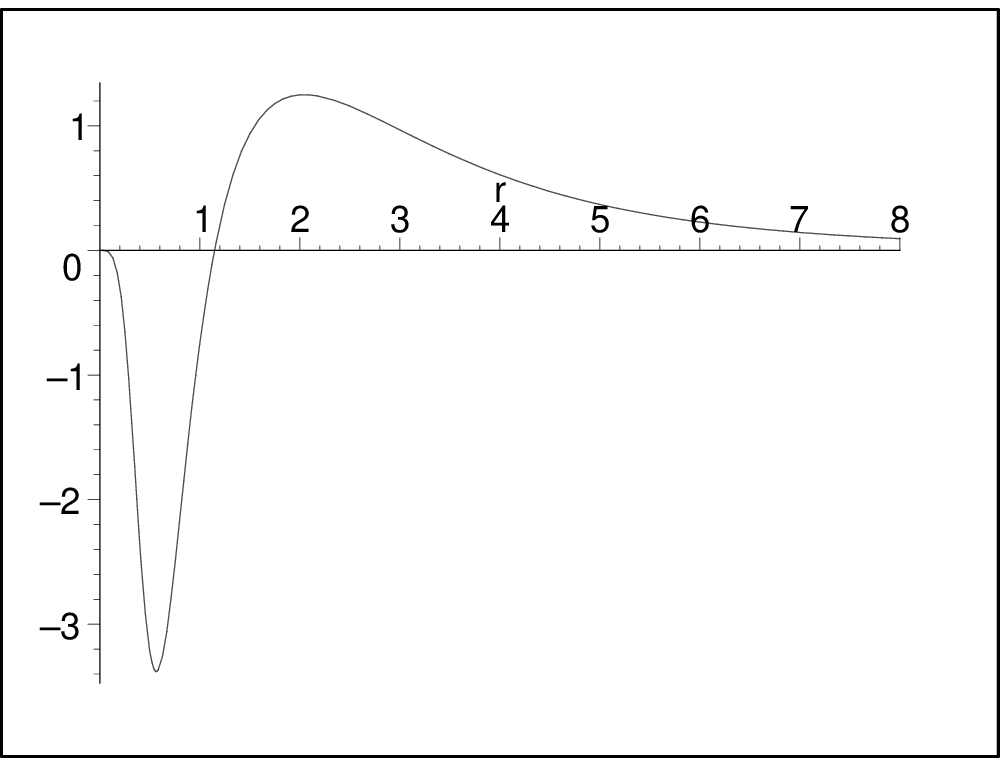, width = 4cm}%
\caption{Energy density associated
with the $\alpha'$ correction for $p=3$, $\rho=l_{s}%
$ and $L=10\,l_{s}$.}\label{lws3}}  
It is a generic feature,
\textit{i.e.,}\ for all values of $p$, $L$ and $\rho$, that this expression dips
into negative values for $r\approx L$. This plays an important role in the
delocalization of the wrapping. In fact, one can now plot the total energy, \ie 
$E+E_{\alpha^{\prime}}$, as a function of the delocalization parameter $L$. We find that
there exists a minimum at finite $L$ (of the order of the string lenght) for a wide range of
the radial parameter $\rho$. 
It should be emphasized that the negative energy
does not point to an instability of the solitonic configuration. One should
see the system as being composed of an infinitely extended brane therefore
possessing an infinite energy. We find that there will always be a minimum for
finite $L$ as long as the correction $E_{\alpha^{\prime}}$ is large enough to
maintain it. One can show that as the typical size of the target space ${\bf S}^{p}$
is increased, the term $E_{\alpha^{\prime}}$ becomes increasingly small taking
the minimum of the total energy to smaller and
smaller values of $L$. Of course, when $E_{\alpha^{\prime}}$ is large ($L$
is then large as well), it is not a correction anymore and there is absolutely no
reason why one should not include stringy corrections of order $\alpha
^{\prime}{}^{2}$ and higher.
There will nevertheless be a region where $\rho$ (the radius of the target manifold) 
is small enough that
$E_{\alpha^{\prime}}$ is in fact a `correction' to $E$ that is significantly
larger than the $\mathcal{O}(\alpha^{\prime}{}^{2})$ corrections. In this
regime, the stabilization of the soliton to a
finite world-volume size will not be affected by the $\mathcal{O}%
(\alpha^{\prime}{}^{2})$ terms. 

The results of this appendix are, strictly speaking, only valid for locally wrapped
D3-branes. As
mentionned above one can hardly justify the use of the spherically symmetric
Atiyah-Manton ansatz as a solution of the equations of motion for $p\neq3$.
With an appropriate ansatz one could probably
generalize this simple approach to other D-branes.


\begin{thebibliography}{99}

\bibitem{skyrme1}
T.~H.~Skyrme, ``A Unified Field Theory Of Mesons And
Baryons,'' Nucl.\ Phys.\ \textbf{31}, 556 (1962).

\bibitem{skyrme2}
T.~H.~Skyrme, ``Meson Theory And Nuclear Matter,''
Proc.\ Roy.\ Soc.\ Lond.\ A \textbf{230}, 277 (1955).

\bibitem{bachas2}
C.~Bachas, M.~R.~Douglas and C.~Schweigert, ``Flux
stabilization of D-branes,'' JHEP \textbf{0005}, 048 (2000)
[arXiv:hep-th/0003037].

\bibitem{callan1}
C.~G.~Callan, C.~Lovelace, C.~R.~Nappi and
S.~A.~Yost, ``Loop Corrections To Superstring Equations Of Motion,''
Nucl.\ Phys.\ B \textbf{308}, 221 (1988).

\bibitem{callan2}
C.~G.~Callan, C.~Lovelace, C.~R.~Nappi and
S.~A.~Yost, ``String Loop Corrections To Beta Functions,'' Nucl.\ Phys.\ B
\textbf{288}, 525 (1987).

\bibitem{leigh}
R.~G.~Leigh, ``Dirac-Born-Infeld Action From Dirichlet
Sigma Model,'' Mod.\ Phys.\ Lett.\ A \textbf{4}, 2767 (1989).

\bibitem{adkins-nappi-witten}
G.S. Adkins, C.R. Nappi and E. Witten, ``Static
Properties Of Nucleons In The Skyrme Model,'' Nucl. Phys. B \textbf{228}, 552 (1983)

\bibitem{corley}
S.~Corley, D.~A.~Lowe and S.~Ramgoolam,
``Einstein-Hilbert action on the brane for the bulk graviton,'' JHEP
\textbf{0107}, 030 (2001) [arXiv:hep-th/0106067].

\bibitem{bachas}
C.~P.~Bachas, P.~Bain and M.~B.~Green, ``Curvature
terms in D-brane actions and their M-theory origin,'' JHEP \textbf{9905}, 011
(1999) [arXiv:hep-th/9903210].

\bibitem{fotopoulos}
A.~Fotopoulos, ``On (alpha')**2 corrections to
the D-brane action for non-geodesic world-volume embeddings,'' JHEP
\textbf{0109}, 005 (2001) [arXiv:hep-th/0104146].

\bibitem{wyllard1}
N.~Wyllard, ``Derivative corrections to the
D-brane Born-Infeld action: Non-geodesic embeddings and the Seiberg-Witten
map,'' JHEP \textbf{0108}, 027 (2001) [arXiv:hep-th/0107185].

\bibitem{wyllard2}
N.~Wyllard, ``Derivative corrections to D-brane
actions with constant background fields,'' Nucl.\ Phys.\ B \textbf{598}, 247
(2001) [arXiv:hep-th/08008125].

\bibitem{horowitz}
G.~T.~Horowitz and L.~Susskind, ``Bosonic M
theory,'' J.\ Math.\ Phys.\ \textbf{42}, 3152 (2001) [arXiv:hep-th/0012037].

\bibitem{strominger}G.~T.~Horowitz and A.~Strominger, ``Black
Strings And P-Branes,'' Nucl.\ Phys.\ B \textbf{360}, 197 (1991).

\bibitem{peet}
A.~W.~Peet, ``TASI lectures on black holes in string
theory,'' arXiv:hep-th/0008241.

\bibitem{polchinski3} J.~Polchinski, ``Dirichlet-Branes and
Ramond-Ramond Charges,'' Phys.\ Rev.\ Lett.\ \textbf{75}, 4724 (1995)
[arXiv:hep-th/9510017].

\bibitem{myers1}
R.~C.~Myers,``Dielectric-branes,''
JHEP {\bf 9912}, 022 (1999)
[arXiv:hep-th/9910053].

\bibitem{GSW}
M.B.~Green, J.H.~Schwarz and E.~Witten, \emph{Superstring
Theory}, Vol. 1 and 2 (Cambridge University Press, 1987).

\bibitem{polchinski2}
J.~Polchinski, \emph{String Theory} Vol. I
and II (Cambridge University Press, 1998).

\bibitem{sen1}
A.~Sen, ``Descent relations among bosonic D-branes,''
Int.\ J.\ Mod.\ Phys.\ A \textbf{14}, 4061 (1999) [arXiv:hep-th/9902105].

\bibitem{sen2}
A.~Sen, ``Stable non-BPS bound states of BPS D-branes,''
JHEP \textbf{9808}, 010 (1998) [arXiv:hep-th/9805019].

\bibitem{bosback1}
O.~DeWolfe, D.~Z.~Freedman, S.~S.~Gubser,
G.~T.~Horowitz and I.~Mitra, ``Stability of AdS(p) x M(q) compactifications
without supersymmetry,'' arXiv:hep-th/0105047.

\bibitem{bosback2}
T.~Shiromizu, D.~Ida, H.~Ochiai and T.~Torii,
``Stability of AdS(p) x S**n x S**(q-n) compactifications,'' Phys.\ Rev.\ D
\textbf{64}, 084025 (2001) [arXiv:hep-th/0106265].

\bibitem{west}
N.~D.~Lambert and P.~C.~West, ``Coset symmetries in
dimensionally reduced bosonic string theory,'' Nucl.\ Phys.\ B \textbf{615},
117 (2001) [arXiv:hep-th/0107209].

\bibitem{gibbons}
G.~W.~Gibbons, G.~T.~Horowitz and P.~K.~Townsend,
``Higher dimensional resolution of dilatonic black hole singularities,''
Class.\ Quant.\ Grav.\ \textbf{12}, 297 (1995) [arXiv:hep-th/9410073].

\bibitem{dobado}
J.~A.~Cembranos, A.~Dobado and A.~L.~Maroto,
``Brane-skyrmions and wrapped states,'' Phys.\ Rev.\ D \textbf{65}, 026005
(2002) [arXiv:hep-ph/0106322].

\bibitem{nak}
M.~Nakahara, \emph{Geometry, Topology and Physics} (Adam
Hilger, 1990).

\end{thebibliography}
\end{document}